\documentclass[a4paper,12pt]{article}

\usepackage[latin1]{inputenc}
\usepackage{float}
\usepackage{amsmath,amssymb}
\usepackage{theorem}
\usepackage{graphicx}

\begin{document}

\title{\bf Unambiguous angular momentum of radiative spacetimes and
asymptotic structure in terms of the center of mass system}

\author{Osvaldo M. Moreschi
\thanks{Email: moreschi@fis.uncor.edu}~
\\
\small FaMAF, Universidad Nacional de C\'{o}rdoba\\ 
\small Ciudad Universitaria, 
(5000) C\'{o}rdoba, Argentina.
}
\date{April 30, 2003}
\maketitle

\begin{abstract}
We present a definition of angular momentum for radiative spacetimes
which does not suffer from any ambiguity of supertranslations. 
We succeed in providing an appropriate notion of {\it intrinsic}
angular momentum; and at the same time a definition of center of mass 
frame at future null infinity.
We use the center of mass frame to present the 
asymptotic structure equations for vacuum spacetimes.

\end{abstract}

PACS numbers: 04.20.-q, 04.20.Cv, 04.20.Ha

\section{Introduction}

The difficulty associated with a sensible definition of angular momentum
at future null infinity $\cal I^+$(scri) is related to the complicated 
structure of the Bondi-Metzner-Sachs (BMS) group of asymptotic symmetries. 
There is a multiplicity of Poincaré subgroups, one for each supertranslation.

Because of this difficulty one can find in the literature
several inequivalent definitions of angular 
momentum\cite{Tamburino66}\cite{Winicour68}\cite{Bramson75}\cite{Prior77}%
\cite{Streubel78}\cite{Penrose82}\cite{Dray84}
which suffer from the so called supertranslation ambiguities.
The only definition of angular momentum without supertranslation
ambiguities is the one we presented in reference \cite{Moreschi86};
however our construction can be criticized on the grounds that we
select a unique frame based on the properties of future
null infinity in the limit for the retarded time going to $-\infty$.
A global choice of this nature can hardly be thought to be able to 
describe the local meaning of intrinsic angular momentum of the
sources at an arbitrary retarded time of a radiating system.

In order to provide with an unambiguous notion of angular momentum in radiative
spacetimes, it is essential to make use of a reference frame system, which
embodies the notion of rest frame\cite{Winicour80}.

The problem of the characterization of the notion of a rest frame 
for isolated systems has been the matter of concerned for many years.
Early attempts used the properties of an optical parameter of a
congruence of null geodesics reaching future null infinity; namely, 
the shear. When the spacetime is stationary, it is known\cite{Newman68}
that one can choose a Bondi system for which the shear $\sigma_0$ is zero;
where the freedom in this choice is just a translation. 
The sections for which the shear is zero are known as `good cuts'.
We see then that for stationary spacetimes there is a simple geometrical way to get
rid off the supertranslation problem. However for radiating spacetimes
things are not so easy. In presence of gravitational radiation, the good cut
equation has in general no solution; but it was suggested in the literature that the 
solution to this can be obtained if $\cal I^+$ is complexified. This approach
has lead to the beautiful construction of $\cal H$ spaces(See reference
\cite{Newman80} and references there in.). Due to the fact that $\cal H$
spaces are intrinsically complex, it remained the difficulty of the
interpretation, in this construction.

This situation motivated us in the past to introduce the so called
`nice sections'\cite{Moreschi88} at $\cal I^+$. These sections are defined
in terms of the nice section equation; which recently has been 
proved\cite{Moreschi98}\cite{Dain00'} to
have solutions in terms of a 4-parameter family of translations. 
Furthermore the solutions have the expected physical properties of
a rest frame\cite{Dain00'}; for example, having a nice section $S_0$,
all other nice sections $S_f$ obtained from future timelike translations
happen to be to the future of $S_0$.

The nice section construction singles out precisely,
in an intrinsic way, a Poincaré structure from the infinite dimensional 
BMS group. In particular, given a fixed observational point $p$ at $\cal I^+$,
there is precisely a 3-degree of freedom of spacelike translations
which generate all the nice sections that contain $p$. 
In contrast, without this constructions there is an infinite dimensional
family of general sections that contain $p$, one for each supertranslation.

The success of the nice section construction allows us to define
a well behaved notion of intrinsic angular momentum along with the
associated notion of center of mass frame.

In section \ref{sec:rest} we briefly review the definition of nice sections.
We make use of charge integrals of the Riemann tensor to present 
in section \ref{sec:intrinsic} both the angular momentum definition and 
the center of mass frame.
In section \ref{sec:asymptotic} we study the asymptotic structure equations
of a general radiating spacetime in terms of the center of mass frame.
We briefly compare our presentation with previous definitions of angular 
momentum and make final comments in section \ref{sec:final}.

We will make use of the GHP\cite{Geroch73} notation.

\section{Rest frame systems at future null infinity}\label{sec:rest}

If a physical system can be ascribed a Poincaré structure, one should be able
to define its momentum vector $P^{\tt a}$ and angular momentum tensor
$J^{\tt ab}$; and then one would define a rest frame by demanding that the 
total momentum has no spacial components in this frame. 
However an isolated gravitating system has the BMS structure\cite{Moreschi87};
therefore it is natural to define a rest frame by transforming to a section 
for which the supermomentum has no spacial components. This brings us to the 
issue of which supermomentum to use. In the nice section construction we 
use\cite{Moreschi88} the supermomentum, that for simplicity we are going to call
`psi'. The supermomentum psi has some interesting properties, but it is important to
emphasize that it does not coincide with the Geroch supermomentum\cite{Geroch77}
or with the Winicour 
one\cite{Tamburino66}\cite{Winicour68}\cite{Winicour80}\cite{Geroch81}.

Given an arbitrary section $S$ of $\cal I^+$, one can, without loss of generality,
choose a Bondi coordinate system $(u,\zeta,\bar\zeta)$, such that $u=0$
determines the section $S$. Then, the supermomentum psi on $S$ is given by the
expression
\begin{equation} \label{supermomentum}
P_{lm}(S)=-\frac{1}{\sqrt{4\pi}} \int_S Y_{lm}(\zeta, \bar \zeta) 
\Psi(u=0,\zeta,\bar \zeta) dS^2, 
\end{equation}
where $dS^2$ is the surface element of the unit sphere on $S$,
$Y_{lm}$ are the spherical harmonics, the scalar $\Psi$ is given by
\begin{equation} \label{Psi}
\Psi \equiv \Psi_2^0 + \sigma_0 \dot{\bar \sigma}_0 +\eth^2 \bar \sigma_0 ,
\end{equation}
where $\Psi_2^0 $ is the leading order asymptotic behavior of the second 
Weyl tensor component,
 $\sigma_0$ is the leading order of  the Bondi shear respectively, where 
we are using $\eth$  to denote the GHP\cite{Geroch73} edth operator 
of the unit sphere, and where a dot means partial derivative with
respect to the retarded time $u$.

The first four components of the supermomentum, namely the case $l=0$ and the
three cases $l=1$, determine the Bondi energy-momentum vector.
More specifically 
\begin{equation} \label{Pbondi} 
\left(P^{\tt a}\right)=\left( 
P_{00},-\frac{1}{\sqrt{6}}(P_{11}-P_{1,-1}),%
\frac{i}{\sqrt{6}}(P_{11}+P_{1,-1}), \frac{1}{\sqrt{3}}P_{10}
\right),
\end{equation} 
where $\tt a=0,1,2,3$.

All other nice sections can be obtained from $S$ by the appropriate 
supertranslations $\gamma$. Therefore one can impose the nice section
condition on $\gamma$. It was shown\cite{Moreschi88}\cite{Moreschi98} 
that this condition can be cast in the following equation
\begin{equation} \label{nice}
\eth^2 \bar \eth^2  \gamma =\Psi(\gamma,\zeta,\bar \zeta) 
 +K^3(\gamma,\zeta, \bar \zeta) M(\gamma),
\end{equation}
where the conformal factor   $K$ can be related to the Bondi momentum by 
\begin{equation} \label{eq:K}
K=\frac{M}{P^{\tt a} l_{\tt a}},
\end{equation}
with
\begin{equation} \label{la}
 (l^{\tt a}) =\left( 1,\frac{\zeta +\bar \zeta }{
1+\zeta \bar{\zeta }},\frac{\zeta -\bar{\zeta }}{i(1+\zeta 
\bar{\zeta )}},\frac{\zeta \bar{\zeta }-1}{1+\zeta \bar{\zeta 
}}\right)
\end{equation}
and $P^{\tt a}$ is evaluated at  the section $u=\gamma$; 
which is calculated through the integral (\ref{Pbondi}).
The rest mass $M$ at the section  is given by
\begin{equation} \label{M}
M=\sqrt{P^{\tt a} P_{\tt a}},
\end{equation}
where the indices are raised and lowered with the Lorentzian flat metric 
$\eta_{ab}$ at scri as described in reference \cite{Moreschi86}.

As it was stated in the introduction, there exists a four parameter family
of solutions of equation (\ref{nice}) with the expected physical 
properties\cite{Moreschi98}\cite{Dain00'}. One can always express
$\gamma(\zeta,\bar\zeta)$ in terms of the translation part 
$\gamma_I(\zeta,\bar\zeta)$ and its proper supertranslation
part $\gamma_{II}(\zeta,\bar\zeta)$; where using spherical
harmonics $\gamma_I$ is expressed in terms of $Y_{00}$ and $Y_{1m}$;
while $\gamma_{II}$ is expressed in terms of $Y_{l_2\, m}$ with
$l_2 \geq 2$. The result is that for each translation $\gamma_I$
there is a proper supertranslation $\gamma_{II}$.

We have assumed that the original section $S$ determined by $u=0$ is a nice
section. In other words $\gamma = 0 $ is already a solution of (\ref{nice}),
and so $K(0,\zeta,\bar\zeta)=1$. For any other nice section $\gamma \neq 0$
in general the Bondi momentum will have spacelike components with respect
to the original system, and so one would have $K \neq 1$. In that case, 
in order to observe that the space
like components of the supermomentum are zero, one needs to make a
Lorentz transformation of the BMS group, that aligns the generator of
time translations with the Bondi momentum, at the new section. 

The need of supertranslations and Lorentz transformations for grasping
the idea of rest frames, is what makes the discussion of radiative
spacetimes so complicated; and it is clear that one single Bondi
system does not provide with a rest frame unless the metric is stationary.

\section{Intrinsic angular momentum and center of mass}\label{sec:intrinsic}

\subsection{Charge integrals of the Riemann tensor}

The rest frames provided by the nice sections construction permit us to
define in an unambiguous way the physical quantities of an isolated system,
in particular its angular momentum.

It is convenient to approach the notion of physical quantities by the
method of charge integrals of the Riemann tensor, as it was used in
our previous work\cite{Moreschi86}.

Given a 2-sphere  $S$ a charge integral of the Riemann tensor
is a quantity of the form:
\begin{equation}
  \label{eq:chargeinte}
  Q_{S} =\int_{S}C 
\end{equation}
where the 2-form $C_{ab}$ is given in terms of the Riemann tensor by
\begin{equation}
  \label{eq:chargeint}
  C_{ab} \equiv R_{ab}^{*\;\;cd}\; w_{cd} ,
\end{equation}
and where the 2-form $w_{ab}$ is to be chosen appropriately.

We will assume that the sphere $S$ is actually a section of $\cal I^+$.
Since the spacetime is asymptotically flat, it is possible to express
the metric, in a neighborhood of scri, around a flat background,
namely
\begin{equation}
  \label{eq:metricexp}
  g_{ab}=\eta_{ab} + h_{ab} ,
\end{equation}
where at scri, the conformal regular metric $\tilde g_{ab}$ is given
by
\begin{equation}
  \label{eq:gtilde1}
  \left. \tilde g_{ab}\right|_{\cal I^+} = \Omega^2 \eta_{ab}  ;
\end{equation}
in other words, at scri one has $\Omega^2 h_{ab} = 0$ .

Let us consider, for a moment, the situation in which the expansion
(\ref{eq:metricexp}) can be extended to the interior of the spacetime;
which, to simplify the discussion, will be assumed to contain no
singularities. Let $\Sigma$ be a spacelike hypersurface in the interior
of the spacetime but that reaches asymptotically future null infinity;
in such a way that in the conformaly completed spacetime, $\Sigma$
can be extended to scri with boundary $S$. Then, it is deduced that
the charge integral on $S$ can be expressed as an integral on $\Sigma$,
by the use of Stokes' theorem, namely
\begin{equation}
  \label{eq:qonsigma}
    Q_{S} =\int_{S}C = \int_{\Sigma} dC .
\end{equation}
The exterior derivative of $C$ can be expressed by
\begin{equation}
  \label{eq:dC}
  dC_{abc} =\frac{1}{3} \epsilon _{abcd} \;^{*}R^{*defg} \;\nabla _{e} w_{fg}
;
\end{equation}
where it is interesting to note that a property of the double dual
of the Riemann tensor is that its trace gives the Einstein tensor,
namely:  ${}^{*}R_{abcd}^{*} g^{bd} =G_{ac} =R_{ac} -\frac{1}{2} g_{ac} R$.
Therefore the previous equation can be written
\begin{equation}\label{eq:dC2}
\begin{split}
dC_{abc}  = & \frac{1}{3} \epsilon _{abcd} \ ^{*}R^{*defg} \ \left(
T_{efg} +\frac{1}{3} g_{ef} v_{g} -\frac{1}{3} g_{eg} v_{f} \right)  \\
= & \frac{1}{3} \epsilon _{abcd} \left( -2G^{dg} v_{g} + {}^{*}R^{*defg}
\ T_{efg} \right) ,
\end{split}
\end{equation}
where $T_{abc}$ is the traceless part of $\nabla_a w_{bc}$ and
$v_c$ its trace; in other words
\begin{equation}
  \label{eq:nablaw}
  \nabla _{a} w_{bc} = T_{abc} +\frac{1}{3} g_{ab} v_{c} -\frac{1}{3}
g_{ac} v_{b} ;
\end{equation}
where it can be checked that $\nabla_a w^{ab}=v^b$.
From equation (\ref{eq:dC2}) one observes that if the vector $v^a$
where a Killing vector of the metric $\eta_{ab}$ and $T_{abc}$
where $O(h)$, then the charge integral will give precisely the
conserved quantities in the context of linearized gravity. This
analysis ensures that this charge integrals admit the appropriate
physical interpretations in the linearized gravity regime.

In what follows, we will only assume that the spacetime is asymptotically
flat at future null infinity\cite{Moreschi87}.

Another interesting property of the double dual of the Riemann tensor is
the one associated with the Bianchi identities, namely 
${}^*R^{*d[efg]} = 0$, from which one can prove the relations\cite{Moreschi86}
\begin{equation}
  \label{eq:R*T}
  {}^*R^{*defg} \, T_{efg} 
=    \frac{2}{3}\; {}^*R^{*defg} \left(T_{(ef)g} - T_{(eg)f} \right) 
;
\end{equation}
where it is important to note that the factor involving the tensor $T_{abc}$
have a very simple form when expressed in terms of the spinorial notation,
that is
\begin{equation}
  \label{eq:Tofw}
\frac{2}{3} \;\left( T_{(ef)g} -T_{(eg)f}
\right) =\nabla _{E' (E} w_{FG)} \;\epsilon _{F' G' } + {\tt c.c.} \;;
\end{equation}
where ${\tt c.c.}$ means complex conjugate.

Observing equation (\ref{eq:dC2}) and recalling the analysis in the
context of linearized gravity, it is natural to study at scri 
the conditions on $w$ given by the equations
\begin{equation}
  \label{eq:divw}
  -\nabla _{A} ^{\;\;B' }\; w^{AB} + {\tt c.c.}= v^{BB'}
\end{equation}
and 
\begin{equation}
  \label{eq:symw}
  \nabla _{E' (E} \; w_{FG)} =0;
\end{equation}
where the vector $v^{BB'} $ is a generator of asymptotic symmetries.
 
The generators of Lorentz rotations are a special case of asymptotic
symmetries. The vector field $v^a$, defined in a neighborhood of scri
is said to be an asymptotic symmetry if it satisfies
\begin{equation}
  \label{eq:asymsym}
  \nabla_{(a} v_{b)} = \frac{S_{ab}}{\Omega} ,
\end{equation}
where $S_{ab}$ has a regular extension to $\cal I^+$.

In general an asymptotic symmetry $v^a$ can be expressed by 
its components, in terms of a null tetrad frame
\begin{equation}
  \label{eq:vasym}
  v^a = v_n \, \ell^a -v_{\bar m}\,  m^a -v_m \, \bar m^a + v_\ell \, n^a ;
\end{equation}
where we are using the standard complex null tetrad that satisfies
$\ell^a\, n_a = - m^a \,\bar m_a =1$ with all other scalar products giving
zero. As usual the tetrad vector $\ell^a$ is chosen to generate a 
congruence of null geodesics reaching scri. A regular tetrad at future
null infinity 
$(\hat\ell^a, \hat m^a, \hat{\bar m}^a, \hat n^a)$
can be constructed from the following relations
\begin{equation}
  \label{eq:lhat}
  \ell^a = \Omega^2 \hat\ell^a ,
\end{equation}
\begin{equation}
  \label{eq:mhat}
  m^a = \Omega \hat m^a ,
\end{equation}
\begin{equation}
  \label{eq:nhat}
  n^a =  \hat n^a ;
\end{equation}
where $\Omega$ can be taken as the inverse of the affine parameter $r$
of the null geodesics reaching scri.

Since the asymptotic symmetries are tangent to $\cal I^+$, the
tetrad components have the following behavior
\begin{equation}
  \label{eq:vn}
  v_n = r v^0_n + v^1_n + O(\frac{1}{r}) ,
\end{equation}
\begin{equation}
  \label{eq:vm}
  v_m = r v^0_m + v^1_m + O(\frac{1}{r}) ,
\end{equation}
\begin{equation}
  \label{eq:vl}
  v_\ell =  v^0_\ell + \frac{v^1_\ell}{r} + O(\frac{1}{r^2}) .
\end{equation}

The leading order behavior of the asymptotic symmetries 
is given by
\begin{equation}
  \label{eq:vm0}
  v^0_{m} =\eth a ,
\end{equation}
\begin{equation}
  \label{eq:vn0}
  v^0_{n} =\frac{1}{2} \left(\eth v_{\bar{m} } 
    + \bar\eth v_{m} \right) =\frac{1}{2}\eth \bar\eth (a+\bar{a} ) ,
\end{equation}
\begin{equation}
  \label{eq:vl0}
  v^0_{l} =\chi (\zeta ,\bar{\zeta } )-u\frac{1}{2} \eth \bar\eth(a+\bar{a} )
\end{equation}
where we are using a Bondi tetrad\cite{Moreschi86}, 
$\chi $ and $a$ are functions on the sphere with spin weight 0
and satisfying: $\chi=\bar\chi$, $\dot{\chi } =0$,  $\dot{a} =0$ and
 $\eth^{2} a=0$.
Let us note that the function ``$a$'' only appears under the action of the
operator edth; then, without loss of generality we can assume that it 
satisfies $\bar\eth \eth a = -a$; which says that $a$ can be expressed
in terms of spherical harmonicas $Y_{lm}$ with $l=1$.

The relation (\ref{eq:divw})  at scri can 
be expressed  in terms of the spinorial components of the regular dyad
\begin{equation}
  \label{eq:wcomp}
w^{AB} =w_{0} \;\hat{\iota}^{A} \hat{\iota}^{B} - w_{1} 
\left(\hat{o}^{A} \hat{\iota}^{B} + \hat{\iota}^{A} \hat{o}^{B} \right)
+ w_{2} \;\hat{o}^{A} \hat{o}^{B}   ,
\end{equation}
by
\begin{equation}
  \label{eq:w2}
  w_{2} =-\frac{1}{3} v_{\bar{m} } ,
\end{equation}
\begin{equation}
  \label{eq:w1}
  w_{1} +\bar{w} _{1} =-\frac{1}{3} v_{\ell} ,
\end{equation}
\begin{equation}
  \label{eq:dotw1}
  \dot{w} _{1} +\dot{\bar{w} } _{1} =-\frac{1}{2} \left(\eth w_{2} 
+\bar\eth \bar{w}_{2} \right) ;
\end{equation}
while condition (\ref{eq:symw}) at scri becomes
\begin{equation}
  \label{eq:ethw2}
  \bar\eth   w_{2} =0 ,
\end{equation}
\begin{equation}
  \label{eq:dotw2}
  \dot{w} _{2} =0 ,
\end{equation}
\begin{equation}
  \label{eq:ethw0}
  \eth w_{0} =-2\;\sigma_0 \;w_{1} ,
\end{equation}
\begin{equation}
  \label{eq:dotw1b}
  \dot{w} _{1} =-\frac{1}{2} \eth w_{2} ,
\end{equation}
\begin{equation}
  \label{eq:dotw0}
  \frac{1}{2} \dot{w} _{0} + \eth w_{1} +\sigma_0 \;w_{2} =0 .
\end{equation}
It is interesting to see whether the solutions of (\ref{eq:divw}) at $\cal I^+$,
namely (\ref{eq:w2})-(\ref{eq:dotw1}) also satisfy
(\ref{eq:ethw2})-(\ref{eq:dotw0}).

Before studying this for the stationary and radiating case separately, let us note
that for any 2-form $w$ defined on a sphere $S$, one can define the charge integral
(\ref{eq:chargeinte}) obtaining
\begin{eqnarray}
  \label{eq:chargegeneral}
  Q_S(w) = 4 \int \left[
- \tilde w_2 (\Psi_1 - \Phi_{10}) + 
  2 \tilde w_1 (\Psi_2 - \Phi_{11} - \Lambda )
- \tilde w_0 (\Psi_3 - \Phi_{21})
\right] dS^2 + {\tt c.c.}
\end{eqnarray}
where $(\tilde w_0, \tilde w_1, \tilde w_2)$ are the components of $w_{AB}$ with
respect to the dyad adapted to $S$, namely for which the vectors $m$ and $\bar m$ 
are tangent to $S$ and $\ell$ is outgoing. If $w$ has a regular extension to scri
we can take the limit of $S$ to a section of $\cal I^+$ obtaining
\begin{eqnarray}
  \label{eq:chargescri}
    Q_S(w) = 4 \int \left[
-  w_2 \Psi_1^0  + 
  2  w_1 \Psi_2^0 
-  w_0 \Psi_3^0 
\right] dS^2 + {\tt c.c.}
\end{eqnarray}
where the upper-script $0$ denotes the leading order behavior of the respective
Weyl components. The charge integral at $\cal I^+$ turns out to be independent of
the Ricci tensor since as we have shown in \cite{Moreschi87} for any
asymptotically flat spacetime, the components of the Ricci tensor go to zero 
faster than the accompanying Weyl components in the above terms. Furthermore, using
that $\Psi_3^0 = -\eth \dot{\bar \sigma}_0$ and equation (\ref{eq:ethw0}) one
can rewrite this  as
\begin{eqnarray}
  \label{eq:chargescri2}
    Q_S(w) = 4 \int \left[
-  w_2 \Psi_1^0  + 
  2  w_1 (\Psi_2^0 + \sigma_0 \dot{\bar \sigma}_0)
\right] dS^2 + {\tt c.c.}
\end{eqnarray}

It is important to note that the solutions of equations (\ref{eq:w2})-(\ref{eq:dotw1})
can always be made to satisfy equations (\ref{eq:ethw2})-(\ref{eq:dotw1b}) and so
one can obtain expression (\ref{eq:chargescri2}) for the charge integral, which
is independent of the validity of equation 
(\ref{eq:dotw0}).

\subsection{Stationary spacetime case}

For the case of stationary spacetimes one can solve the set of equations
(\ref{eq:ethw2})-(\ref{eq:dotw0}) with solution
\begin{align}
 w_{2} &=-\frac{1}{3} \bar\eth \bar a  , \label{eq:wsolut1} \\
 w_{1} &=w_{1}^{00} ( \zeta ,\bar{\zeta },\sigma_0  )+
            \frac{1}{6}\,u\, \eth \bar\eth \bar a  , \label{eq:wsolut2}\\
 w_{0} &=w_{0}^{00} (\zeta ,\bar{\zeta }, \sigma_0 ) +u\left( -2 \eth w_{1}^{00} 
            + \frac{2}{3} \sigma_0  \bar\eth \bar a 
          \right) - \frac{1}{6}\, u^2 \, \eth^{2} \bar\eth \bar a  \label{eq:wsolut3}
\end{align}
where $\bar\eth^2 \bar a =0$,
$w_{1}^{00} $ 
and $w_{0}^{00}$ are spin weight 0 and 1 functions 
respectively that solve the equations
\begin{equation}
  \label{eq:w100}
\eth ^{2} w_{1}^{00} =\frac{1}{3}\eth \sigma_0 \;\bar{\eth}\bar{a} +
\frac{1}{2} \sigma_0\;\eth \bar{\eth}\bar{a} =
-\eth\sigma_0 \;w_{2} -\frac{3}{2} \sigma_0 \;\eth w_{2} 
\end{equation}
and
\begin{equation}
  \label{eq:w000}
\eth  w_{0}^{00} =-2\sigma_0 w_{1}^{00} .
\end{equation}
Let us note that if one uses the potential $\delta$ of the shear satisfying
\begin{equation}
  \label{eq:alfasigma}
  \sigma_0 = \eth^2 \delta ,
\end{equation}
then, the component $w_1$ can be expressed by
\begin{equation}
  \label{eq:w1alfa}
  w_1 = b + \frac{1}{3} \eth \delta \bar\eth \bar a 
+ \frac{1}{6} (u - \delta) \eth \bar\eth \bar a ;
\end{equation}
where the spin 0 quantity $b$ satisfies $ \dot b =0$ and $\eth^2 b = 0$.

This procedure provides with
a two-form with the functional dependence
\begin{equation}
  \label{eq:w0}
  w_{AB}^{0} =w_{AB}^{0} \left( 
u,\zeta ,\bar{\zeta } 
;
\sigma_0(\zeta ,\bar{\zeta } ),a,b
\right) .
\end{equation}

Let us observe that from one of the Bianchi identities at scri, it is obtained
\begin{equation}
  \label{eq:ethpsi2}
  \eth \Psi_2^0 = \dot \Psi_1^0 - 2\, \Psi_3^0 \sigma_0 ,
\end{equation}
but for a stationary spacetime $\dot \Psi_1^0 = 0$ and also
$ \Psi_3^0 = 0$
in a Bondi system; therefore, one deduces that $\Psi_2^0$ is a constant.
Then, since the nice section condition requires 
$\Psi =  \Psi_2^0 + \sigma_0 \dot{\bar{\sigma}}_0 +\eth^2 \bar \sigma_0 = %
 \Psi_2^0 +\eth^2 \bar \sigma_0 $ to be a constant; one concludes that
$\eth^2 \bar \sigma_0 =0 $ and therefore that $\sigma_0 = 0$.
In other words the nice section condition coincides in the stationary
case with the good cut equation condition. However, as we have noticed,
the nice section equation has solutions in the radiating case,
while the good cut equation does not.

In the stationary case we have that the charge integrals of the Riemann tensor
at scri are given by
\begin{equation}
  \label{eq:chargestat}
Q_S(w)=  \int_{S}C   =4\int_{S}\Bigl( -w_{2} \, \Psi_{1}^{0} 
+ 
 2 \,w_{1} \,  \Psi _{2}^{0} 
\Bigr) \;dS^{2} + {\tt c.c.} 
\, ;
\end{equation}
where $w_2$ is given by (\ref{eq:wsolut1}) and $w_1$ is
\begin{equation}
  \label{eq:w1alfa0}
  w_1 =  b 
+ \frac{1}{6} u  \eth \bar\eth \bar a .
\end{equation}
Let us note that $a$ involves 6 real constants associated with the
Lorentz rotations, and that since $\Psi_2^0$ is a real quantity, $b$
contributes to the charge integral $Q_S(w)$ with other four real constants
associated with translations.

We see then that this construction provides with the appropriate rest
frames for stationary spacetimes; and that the vector $v^a(a,b+\bar b)$ 
calculated
from the divergence of the 2-form $w$, equation (\ref{eq:divw}), is
a generator of BMS symmetries.

The first term in the integrand of equation (\ref{eq:chargestat}) includes
the Weyl component $\Psi_1$ which is known to describe the angular
momentum in the Kerr geometry for example. In the second term
we recognize the component $\Psi_2^0$ which form the
supermomentum psi for this particular
stationary case. Therefore equation (\ref{eq:chargestat})
has the same structure as the expression that, in special relativity,
relates the angular momentum $J^{\tt a  b}$, 
the intrinsic angular momentum $S^{\tt a b}$ and
the linear momentum $P^{\tt a}$, namely
\begin{equation}
  \label{eq:j}
  J^{\tt a  b}=S^{\tt a b}+R^{\tt a} \, P^{\tt b} - P^{\tt a}\, R^{\tt b} ,
\end{equation}
where ${\tt a,b}$ are numeric spacetime indices. Given a rest reference
frame in Minkowski spacetime one needs to use the spacelike translation
freedom $(R^{\tt b})$ appearing in expression (\ref{eq:j}) in order to single out
the center of mass reference frame. In the center of mass frame
one has $ J^{\tt a  b}=S^{\tt a b}$; that is the total momentum 
coincide with the intrinsic angular momentum. Since the intrinsic
angular momentum satisfies $S^{\tt ab} P_{\tt b}=0$, one can characterize
the center of mass frame as that rest frame for which 
$J^{0i}=0$; where $i=1,2,3$ is a spacelike numeric index.

In the nice section construction we have seen that given an observation
point $p$ at scri, there is a 3-family of rest frames, associated with
the choice of spacelike translations. We also observe that the charge
integral (\ref{eq:chargestat}) has the appropriate angular momentum
behavior. Therefore we also need a condition analogous to the
Minkowskian case to single out the center of mass reference frames.

It can be seen that the condition 
to be imposed in the charge integral case is
\begin{equation}
  \label{eq:acondit}
  Q_S(a)=0 \qquad \text{for all}\quad a=\bar a ;
\end{equation}
where it is understood that one takes $b=0$ in this equation.
The quantity $a$ is in principle complex; so this condition
makes use precisely of a 3-degree of freedom, which is 
associated with spacelike translations.

This is the appropriate condition that leaves a one-dimensional family of
nice sections $S$ that can legitimately be called 
{\em center of mass frames}.
We can see that the center of mass frames  are generated by time translations
in the nice section construction.

Using these frames $S_{\tt cm}$, the intrinsic angular momentum $j$ is defined
through
\begin{equation}
  \label{eq:instrin1}
  j(w)=Q_{S_{\tt cm}}(w) ;
\end{equation}
where to determine $w$ one chooses $a= -\bar a$ and $b=0$.

The same charge integral can also be used to calculate the Bondi momentum,
which requires to take $a=0$ and $b\neq0$.

This prescription singles out the center of mass frame for stationary
spacetimes and a Poincaré subgroup of BMS generators.

\subsection{Radiating spacetime case}

In the presence of gravitational radiation, the shear $\sigma_0$ depends
on the time $u$ and the set of equations (\ref{eq:w2})-(\ref{eq:dotw0}) in general
have no solution. 
But it is interesting to note that one can define a 2-form $w$ at scri, 
which in some sense is the closest to the stationary prescription, 
as we describe next.

Let $(u,\zeta.\bar\zeta)$ be some Bondi coordinate system at future null infinity.
A generator of scri can be identified with the line determined by $\zeta={\tt constant}$
($\bar\zeta={\tt constant}$).
Let us consider a point along a particular generator of scri, denoted by $p(\tau)$,
with $\tau$ a monotonically increasing time parameter.

Let us recall\cite{Dain00'} that
the set of nice sections form a four parameter $(T,\vec R)$
family  that we now label $S_{(T,\vec R)}$; 
where  $(T,\vec R)$ can be identified with a translation of the BMS group.

Then,
for a given fixed $\tau$, one has a 3-parameter $(\vec R)$ family of nice sections 
$S_{(T,\vec R)}$ that contain the point $p(\tau)$; where all this family share the same
$T(\tau)$.
Given one of this nice sections $S_{(T,\vec R)}$, we can always identify it
with the condition $u=\gamma_S$, where $\gamma_S(\zeta,\bar\zeta)$ is the supertranslation
that defines the corresponding section. On $S_{(T,\vec R)}$ we define the 2-form
$w_{S_{(T,\vec R)}}(u=\gamma_S,\zeta,\bar\zeta; a,b)$ as the solution on 
$S_{(T,\vec R)}$ of equations (\ref{eq:divw}) and (\ref{eq:symw}) 
for the radiation data 
$\sigma_{S_{(T,\vec R)}}(u,\zeta,\bar\zeta) =  \sigma_0(\gamma_S,\zeta,\bar\zeta)$.
Since by definition $\sigma_{S_{(T,\vec R)}}$ does not depend on the time
coordinate $u$, the 2-form $w_{S_{(T,\vec R)}}$ is the restriction on $S_{(T,\vec R)}$ 
of the solution of the stationary problem when the radiation data is
$\sigma_{S_{(T,\vec R)}}$. 
More concretely, one can express the 2-form $w_{S_{(T,\vec R)}}$ in terms of its
spinor components by the prescription stated in equations 
(\ref{eq:wsolut1})-(\ref{eq:w000}), if we chose the Bondi system such that
$S_{(T,\vec R)}$ is characterized by $u=\tt constant$.

In order to single out the center of mass section $S_{cm}$ from the 3-parameter
$(\vec R)$ family of nice sections that contain the point $p(\tau)$, we 
demand 
\begin{equation}
  \label{eq:acondit2}
  Q_{S_{(T,\vec R)}}(a)=0 \qquad \text{for all}\quad a=\bar a .
\end{equation}
This is the analog in the Minkowskian case of the condition $J^{0i}=0$; that
we mentioned earlier.
Let us note that the quantity $a$ involves six real parameters; so that the
condition $a=\bar a$ involves three degrees of freedom which are taking care
of by the corresponding freedom involved in $(\vec R)$.
In this way, for each $\tau$ we select the unique $S_{cm}(\tau)$ which contains
the observation point $p(\tau)$ at scri.

Given the radiation data $\sigma_0(u,\zeta,\bar \zeta)$, 
this construction provides with a 
smooth\cite{Dain00'}  one-parameter family of non-crossing 
center of mass sections $S_{cm}(\tau)$ along with a smooth 2-form
\begin{equation}\label{eq:wrad}
w(\tau,\zeta,\bar\zeta; a,b)=
w_{S_\tau}(u=\gamma_{S_\tau},\zeta,\bar\zeta; a,b)
\end{equation}
on scri.

The 2-form $w_{AB}$ as defined by (\ref{eq:wrad}) will not satisfy
all equations (\ref{eq:w2})-(\ref{eq:dotw0}); more concretely, most of
the equations involving time derivatives will not be satisfied in general, but
the error will be $O(\dot\sigma_0)$. That is, if $\dot\sigma_0 =0$
on a particular center of mass section $S_{cm}$, then 
  all equations  (\ref{eq:divw}) and (\ref{eq:symw}),
or equivalently (\ref{eq:w2})-(\ref{eq:dotw0}),
 are satisfied
on $S_{cm}$.

Using these center of mass frames $S_{\tt cm}$, the intrinsic angular momentum 
$j$ is defined through
\begin{equation}
  \label{eq:instrin}
  j(w)=Q_{S_{\tt cm}}(w) ;
\end{equation}
where as before, in order to pick up the intrinsic angular momentum, one
takes $a=-\bar a$ and $b=0$.

With this definition the  charge integral of the Riemann on $S_{cm}$, can be expressed by
\begin{equation}
  \label{eq:charge}
\begin{split}
Q_{S_{cm}}(w)=  \int_{S_{cm}}C   =4\int_{S_{cm}}\Bigl( -& \, w_{2} \left( \Psi_{1}^{0} 
+ 2\sigma_0 \eth\bar{\sigma_0 } +\eth\left( \sigma_0 \bar{\sigma_0 } \right)
\right)  \Bigr. \\
+ \Bigl.
& \, 2 \,w_{1} \left( \Psi _{2}^{0} +\sigma_0 \dot{\bar{\sigma_0 } } +
\eth^{2} \bar{\sigma_0 } \right) \Bigr) \;dS^{2} + {\tt c.c.} 
\end{split}
\, ;
\end{equation}
where we have used that
\begin{equation}
  \label{eq:w1ethsigma}
\int\limits_{S_{cm}}\left( w_{1}\;\eth^2 \bar{\sigma_0 } \right) \ dS^{2}  
=
\int\limits_{S_{cm}}\left( \sigma_0 \, \eth\bar{\sigma_0 } 
  +\frac{1}{2} \eth \left(
\sigma_0 \bar{\sigma_0 } \right) \ \right) w_{2} \ dS^{2} .
\end{equation}

It should be emphasized that in this way, given an 
asymptotically flat spacetime, one constructs a unique regular family of 
non intersecting sections\cite{Dain00'} at future null
infinity, which are the center of mass frames, and that can be
used to describe the detailed multipolar asymptotic structure of the spacetime.
 
\begin{figure}[htbp]
\centering
\includegraphics[clip,width=0.6\textwidth]{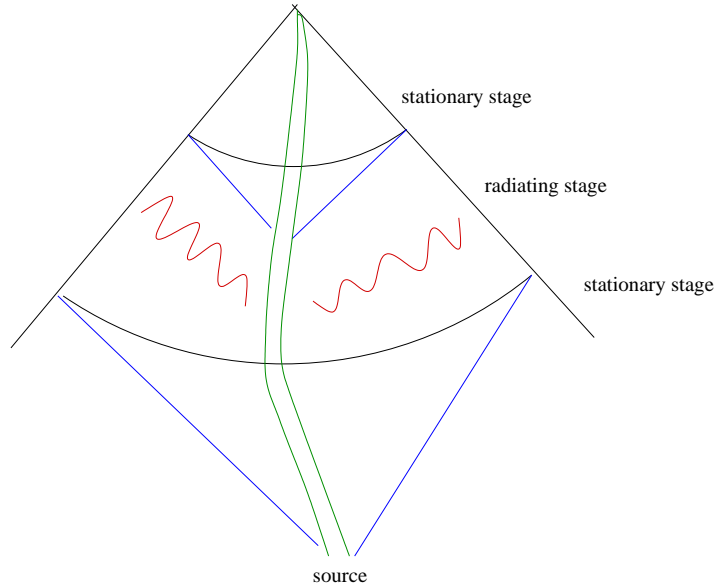}
\caption{Behavior of the notion of intrinsic angular momentum
in a spacetime with three stages; beginning with a stationary stage,
continuing with a radiating stage and ending with a stationary stage.
}
\label{fig:threestages}
\end{figure}

In order to see whether our definition of angular momentum 
has the appropriate behavior in the presence of gravitational radiation,
let us imagine a system 
in which one can distinguish three stages; starting with a stationary
regime, passing through a radiating stage and ending in a stationary regime
as depicted in figure \ref{fig:threestages}.
By construction it is clear that our definition gives the correct
intrinsic angular
momentum in the first and third stage; even though the center of mass frames
of the first and third stage are related in general by a supertranslation.

This is the only definition of angular momentum that satisfies 
this expected property.

\section{Asymptotic structure of asymptotically flat spacetimes}\label{sec:asymptotic}

\subsection{Basic variables}

Having obtained the center of mass frames, we now take the opportunity to 
study the asymptotic structure in terms of a null tetrad which is adapted
to the center of mass frames.

We start by defining the basic variables that we will use.

One can express the asymptotic geometry in terms of a complex null tetrad
$\left( \ell ^{a},m^{a},\overline{m}^{a},n^{a}\right) $ with the properties:
\begin{equation}\label{eq:produc}
g_{ab}\;\ell ^{a}\;n^{b}=-g_{ab}\;m^{a}\;\overline{m}^{b}=1
\end{equation}
and all other possible scalar products being zero, the metric can be
expressed by
\begin{equation}
g_{ab}=\ell _{a}\;n_{b}+n_{a}\;\ell _{b}-m_{a}\;\overline{m}_{b}-\overline{m}%
_{a}\;m_{b}.
\end{equation}
Newman and Penrose\cite{Newman62'} showed that such a null tetrad is 
easily related to a dyad of spinors $(o^A, \iota^A)$, and that they are 
very useful for the discussion of the asymptotic structure.

Using the null polar coordinate system $(x^{0},x^{1},x^{2},x^{3})=\left( u,r,%
(\zeta +\overline{\zeta}),\frac{1}{i}(\zeta -\overline{\zeta})\right)$
one can express the null tetrad as:
\begin{equation}
\ell _{a}=\left( du\right)_{a} 
\label{uno}
\end{equation}
\begin{equation}
\ell ^{a}=\left( \frac{\partial }{\partial \,r}\right) ^{a} 
\label{dos}
\end{equation}
\begin{equation}
m^{a}=\xi ^{i}\left( \frac{\partial }{\partial x^{i}}\right) ^{a} 
\end{equation}
\begin{equation}
\overline{m}^{a}=\overline{\xi}^{i}\left( \frac{\partial }{\partial x^{i}}\right) ^{a} 
\label{tres}
\end{equation}
\begin{equation}\label{eq:vecn}
n^{a}=\,\left(\frac{\partial}{\partial \,u} \right)^{a}
+ \,U\,\left( \frac{\partial }{\partial \,r}\right)^{a}
+ X^{i}\,\left(\frac{\partial }{\partial \,x^{i}}\right)^{a} 
\end{equation}
with $i=2,3$  and where the components $\xi^{i}$, $U$ and $X^{i}$ are:
\begin{equation}
\xi ^{2}=\frac{\xi _{0}^{2}}{r} + O\left(\frac{1}{r^2}\right),
\qquad \xi ^{3}=\frac{\xi_{0}^{3}}{r} + O\left(\frac{1}{r^2}\right) , 
\end{equation}
with
\begin{equation}\label{eq:xileading}
\xi _{0}^{2}=\sqrt{2}P_{0}\;V,\qquad \xi _{0}^{3}=-i\xi _{0}^{2},
\end{equation}
where $V=V(u,\zeta,\bar\zeta)$ and
the square of $P_0 = \frac{(1+\zeta \bar\zeta)}{2}$ is the conformal factor 
of the unit sphere;
\begin{equation}
U=rU_{00}+U_{0}+\frac{U_{1}}{r} +  O\left(\frac{1}{r^2}\right), 
\end{equation}
where
\begin{equation}
U_{00}=\frac{\dot{V}}{V},\quad U_{0}=-\frac{1}{2}K_{V},\quad U_{1}=-\frac{%
\Psi _{2}^{0}+\overline{\Psi}_{2}^{0}}{2}, 
\end{equation}
where $K_V$ is the  curvature of the 2-metric
\begin{equation}\label{eq:desphere}
dS^{2}=\frac{1}{V^2 \, P_0^{2}}\;d\zeta \;d\bar{\zeta} ;
\end{equation}
where the regular conformal metric restricted to scri is precisely
$\tilde g \mid_{\cal I^+} = - dS^2$. In terms of the edth operator
$\eth_V$ 
of the sphere (\ref{eq:desphere}) the curvature $K_V$ is given by
\begin{equation}
K_{V}=\frac{2}{V}~\overline{\eth }_{V}\eth _{V}\, V-\frac{2}{V^{2}}~\eth _{V}V~%
\overline{\eth }_{V}V+V^{2} .
\end{equation}
Finally, the other components of the vector $n^a$ have the asymptotic form
\begin{equation}
 X^{2}= O\left(\frac{1}{r^2}\right) ,
\qquad X^{3}= O\left(\frac{1}{r^2}\right)  .
\end{equation}

\subsection{Asymptotic behavior of the spin coefficients}

The optical spin coefficients of the outgoing congruence of null geodesics are
\begin{equation}
\rho= -\frac{1}{r} - \frac{\sigma_0 \bar\sigma_0}{r^3} + O(\frac{1}{r^4})
\end{equation}
and
\begin{equation}
\sigma= \frac{\sigma_0}{r^2} 
+ \frac{\sigma_0^2 \bar\sigma_0 - \frac{1}{2}\Psi_0^0}{r^4}
+ O(\frac{1}{r^5})
;
\end{equation}
while their primed version, associated to the vector field $n^a$, are
\begin{equation}
\rho'= -\frac{U_0}{r} 
+ \frac{\Psi_2^0 - \sigma_0 \sigma'_0 + \eth_V \bar\tau_0}{r^2}
+ O(\frac{1}{r^3})
\end{equation}
and
\begin{equation}
\sigma'= \frac{\sigma'_0}{r} + O(\frac{1}{r^2})
;
\end{equation}
where
\begin{equation}
\tau_0 = \bar\eth_V \sigma_0
\end{equation}
and
\begin{equation}
\sigma'_0 = \frac{\dot V}{V} \bar \sigma_0 - \dot{\bar \sigma}_0 
.
\end{equation}

Let us also note that
\begin{equation}
  \label{eq:epsilonp1}
  \epsilon' = \epsilon'_0 + \left(\frac{1}{r}\right) ;
\end{equation}
with
\begin{equation}
  \label{eq:epsilon0p}
  \epsilon'_0 = \frac{\dot V}{2\, V} .
\end{equation}


\subsection{Asymptotic behavior of the Weyl components}
The leading order behavior of the Weyl components $\Psi_2$, $\Psi_3$ and 
$\Psi_4$ satisfy the following relations
\begin{equation}
\Psi_2^0 + \sigma_0 \, \dot{\bar\sigma}_0 + \eth_V^2 \bar\sigma_0    
= \bar\Psi_2^0  +  \bar\sigma_0 \, \dot \sigma_0 + \bar\eth_V^2 \sigma_0 
,
\end{equation}
\begin{equation}
\begin{split}
\Psi_3^0= & \eth_V \sigma'_0 - \bar\eth_V \rho'_0 \\
=& -\eth_V \dot{\bar\sigma}_0 + \frac{\dot V}{V} \eth_V \bar\sigma_0 
   + \left( \frac{\eth_V \dot V}{V} -\frac{\dot V \eth_V V}{V^2} \right)
   \bar\sigma_0 -\frac{1}{2} \bar\eth_V K_V 
\end{split}
\end{equation}
and
\begin{equation}
\begin{split}
\Psi_4^0  = & \dot \sigma'_0 - 2 U_{00} \sigma'_0 - \bar\eth_V^2 U_{00} \\
= 
&
- \ddot{\bar{\sigma}}_0
+ 3 \frac{\dot V }{V}\dot{\bar{\sigma}}_0
+  \left( \frac{\ddot V }{V}
  - 3 \frac{\dot V^2 }{V^2} \right) \bar\sigma_0 \\
& -\frac{\bar\eth_V^2 \dot V}{V} 
+ 2 \frac{\bar\eth_V \dot V \, \bar\eth_V V}{V^2}
- 2 \frac{\dot V \, (\bar\eth_V V)^2}{V^3}
+  \frac{\dot V \, \bar\eth_V^2 V}{V^2}
.
\end{split}
\end{equation}

\subsection{The asymptotic gauge freedom}
\subsubsection{Coordinate and tetrad transformations}
Let us consider the main gauge freedom admitted in our calculation 
which is of the form
\begin{align}
\tilde u &= \alpha(u,\zeta,\bar\zeta) + 
 \frac{\tilde u_1(u,\zeta,\bar\zeta)}{r}  \label{eq:tildeu}
+ O\left(\frac{1}{r^2}\right) ,\\
\tilde r &= \frac{r}{w(u,\zeta,\bar\zeta)}  \label{eq:tilder}
+ O\left(r^0\right)  ,\\
\tilde \zeta & = \zeta + O\left(\frac{1}{r}\right) .\label{eq:tildezeta}
\end{align}
with $\dot \alpha >0$.
The possible further transformation of the coordinates of the sphere
$(\zeta,\bar\zeta)$ into itself, will complicate the discussion 
unnecessarily.

The condition $g^{\tilde u \tilde r}=1$ imposes the relation
\begin{equation}
w = \dot\alpha .
\end{equation}

This asymptotic coordinate transformation is associated to a corresponding
null tetrad transformation; which in the leading orders is given by
\begin{equation}
\begin{split}
\tilde \ell & = d\tilde u = \dot\alpha \, du
+ \alpha_\zeta  \, d\zeta + \alpha_{\bar\zeta}  \, d\bar\zeta 
+O\left(\frac{1}{r}\right) \\
&=  \dot\alpha  \, \ell
- \frac{\eth_{V} \alpha}{r}  \, \bar m 
- \frac{\bar\eth_V \alpha}{r}  \, m 
+O\left(\frac{1}{r}\right) ,
\end{split}
\end{equation}
\begin{equation}
\begin{split}
\tilde n &= 
 \frac{\partial}{\partial \tilde u}  + O\left( \frac{1}{r}\right)
 =  \frac{1}{\dot\alpha}  \,  \frac{\partial}{\partial u}
+ O\left( \frac{1}{r}\right)\\
&=  \frac{1}{\dot\alpha}  \, n
+ O\left( \frac{1}{r}\right) ,
\end{split}
\end{equation}
\begin{equation}
\begin{split}
\tilde m &= 
\frac{\sqrt{2}\tilde P}{\tilde r}  \frac{\partial}{\partial \tilde \zeta}  
  + O\left( \frac{1}{r^2}\right)
 = \frac{\sqrt{2}P_0 \tilde V\, w }{ r}  
\left( -\frac{\alpha_\zeta}{\dot\alpha} 
\,  \frac{\partial}{\partial u}
+ \frac{\partial}{\partial \zeta} \right)
+ O\left( \frac{1}{r^2}\right)\\
&=  -\frac{\sqrt{2}P_0 \tilde V\, w }{ r}  
\frac{\alpha_\zeta}{\dot\alpha} \,  n
+ \frac{\tilde V \, w}{V}m
+ O\left( \frac{1}{r^2}\right) ;
\end{split}
\end{equation}
since the metric expressed in terms of the new null tetrad 
must coincide with the metric
expressed in terms of the original null tetrad, it is deduced that
\begin{equation}\label{eq:tildeV}
\tilde V = \frac{V}{w}=\frac{V}{\dot\alpha} ;
\end{equation}
therefore
\begin{equation}
\tilde m = m
  -\frac{\eth_V \alpha}{r \, \dot\alpha } \,  n
+ O\left( \frac{1}{r^2}\right) .
\end{equation}

The null tetrad transformation equations can be used to write the leading order
transformation relations for the spinor dyad associated to the 
null tetrad\cite{Geroch73}; namely
\begin{equation}
\tilde o^A = \sqrt{\dot\alpha} 
\left( o^A - \frac{\eth_{V} \alpha}{r\, \dot\alpha}\, \iota^A \right)
\end{equation}
and
\begin{equation}
\tilde \iota^A = \frac{1}{\sqrt{\dot\alpha}}\, \iota^A .
\end{equation}
Taking into account higher order transformations would include an equation of 
the form
\begin{equation}
  \label{eq:iotagenral}
  \tilde \iota^A = \frac{1}{\sqrt{\dot\alpha}}
  \left( \iota^A + h\, o^A \right);
\end{equation}
where in principle $h$ could be of order $O\left(r^0\right)$.

Analogous equations to (\ref{eq:lhat})-(\ref{eq:nhat}) relate the 
regular dyad at future null infinity with the spacetime ones, namely
\begin{equation}
  \label{eq:omicronhat}
  \hat o^A = \Omega^{-1}\, o^A ,
\end{equation}
\begin{equation}
  \label{eq:iotahat}
  \hat \iota^A =  \iota^A .
\end{equation}

Then, the regular dyad at future null infinity is given by
\begin{equation}
\hat{\tilde o}^A = \tilde\Omega^{-1}\, \tilde o^A = 
\frac{r}{w}\sqrt{\dot\alpha} 
\left( o^A - \frac{\eth_{V} \alpha}{r\, \dot\alpha}\, \iota^A \right)
=\frac{1}{\sqrt{\dot\alpha}}
\left(\hat o^A - \frac{\eth_{V} \alpha}{\dot\alpha}\, 
\hat\iota^A \right)
,
\end{equation}

\begin{equation}
\hat{\tilde \iota}^A = \tilde \iota^A = \frac{1}{\sqrt{\dot\alpha}}\, 
\hat \iota^A ;
\end{equation}
where we are using $\Omega = \frac{1}{r}$.

\subsubsection{Transformation of $\Psi_2^0$ and $\Psi_1^0$}
We can now easily calculate the component $\Psi_2$ of the Weyl tensor, 
in leading orders,
with respect to the new null tetrad, obtaining
\begin{equation}\label{eq:tpsi2}
\tilde \Psi_2^0 = \tilde{\Omega}^{-1} \Psi_{ABCD}
\hat{\tilde o}^A \hat{\tilde o}^B \hat{\tilde \iota}^C \hat{\tilde \iota}^D
= \frac{1}{\dot\alpha^3} 
\left( \Psi_2^0 - 2\frac{\eth_V\, \alpha}{\dot\alpha} \Psi_3^0
+ \frac{(\eth_V \alpha)^2}{\dot\alpha^2}\, \Psi_4^0 \right) .
\end{equation}
Similarly for $\Psi_1^0$ one has
\begin{equation}\label{eq:tpsi1}
\tilde \Psi_1^0 =  
\frac{1}{\dot\alpha^3} 
\left( 
\Psi_1^0 - 3 \frac{\eth_V\, \alpha}{\dot\alpha} \Psi_2^0 
+ 3 \frac{(\eth_V\, \alpha)^2}{\dot\alpha^2} \Psi_3^0
- \frac{(\eth_V \alpha)^3}{\dot\alpha^3}\, \Psi_4^0 
\right) .
\end{equation}

\subsubsection{Transformation of some spin coefficients }
Recalling that the shear is defined by
\begin{equation}
\sigma = o^A \bar\iota^{A'}\, o^B \nabla_{AA'} o_B 
= \frac{\sigma_0}{r^2} +O\left(\frac{1}{r^3}\right);
\end{equation}
then, using
\begin{equation}
\Lambda =  - \frac{\eth_{V} \alpha}{r\, \dot\alpha} 
\end{equation}
one obtains, for the leading orders behavior of the new shear
\begin{equation}
\begin{split}
\tilde\sigma 
=&
\sqrt{\dot\alpha} 
\left( o^A + \Lambda \, \iota^A \right)
\left( \iota^{A'} + h \, o^{A'} \right)
\left( o^B + \Lambda \, \iota^B \right) \\
&\nabla_{AA'}
\sqrt{\dot\alpha} 
\left( o_B + \Lambda \, \iota_B \right) +O\left(\frac{1}{r^3}\right)
\\
=&
\dot\alpha
\left(\sigma - \eth_m \Lambda 
- \Lambda \mid\!\supset'   \Lambda 
\right. 
 \left.
 +\Lambda \tau - \Lambda^2 \rho' -  \Lambda^3 \kappa' 
\right) +O\left(\frac{1}{r^3}\right)
\end{split}
\end{equation}
where we have used $\mid\!\supset'$ to denote the thorn primed operator,
$\kappa=0$, 
and where $\Lambda$ is recognized as a quantity of \{$p,q$\}\cite{Geroch73} type
\{2,0\}.
This equation implies that
\begin{equation}
\begin{split}
  \label{eq:tsigma0}
  \tilde\sigma_0 
= 
 \frac{1}{\dot\alpha}
\left(\sigma_0 
+ \frac{1}{\dot \alpha} \eth_V^2 \alpha 
+ \frac{1}{\dot \alpha^3 }
 \ddot \alpha \left(\eth_V \alpha \right)^2
  - \frac{2}{\dot \alpha^2} \eth_V \alpha \; \eth_V \dot \alpha
  - \frac{2}{\dot \alpha^2}\left(\eth_V \alpha \right)^2
    \frac{\dot V}{V}
\right)
\end{split}
\end{equation}

In order to calculate the transformation of other spin coefficients, as
for example $\rho'$, $\sigma'$ and $\tau$, one would need to determine
the quantity $h$.
However since the leading order behavior of these spin coefficients can
be put in terms of $V$ and $\sigma_0$, one can calculate its
transformations from those of the last two quantities.

\subsubsection{Transformation of the edth operator}
Let us recall that the edth operator can be defined in terms of the
conformal factor $P=V\, P_0$ of the sphere, by the relations
\begin{equation}
  \label{eq:ethoper}
  \eth_V f = \sqrt{2} P^{1-s} \frac{\partial}{\partial \zeta}\left(P^s f\right)
\end{equation}
and 
\begin{equation}
  \label{eq:ethboper}
  \bar\eth_V f = \sqrt{2} P^{1+s} \frac{\partial}{\partial \bar\zeta}
  \left(P^{-s} f\right) ;
\end{equation}
where $s$ is the spin weight of the quantity $f$.

Then one can prove that the transformed edth operator
acts as 
\begin{equation}
  \label{eq:ethtilde}
  \tilde{\eth}_{\tilde V} f 
= \frac{\tilde V}{V}
\left( 
  \eth_V f - \frac{\eth_V \alpha}{\dot \alpha} \dot f
\right)
+ s \, f 
\left[
  \eth_V\left(\frac{\tilde V}{V}\right) - 
  \frac{\eth_V \alpha}{V\, \dot\alpha} \dot{\tilde V}
\right] .
\end{equation}

\subsubsection{Transformations to a Bondi system}\label{sec:trbondi}
If the target coordinate system is a Bondi system, then one has
\begin{equation}
  \label{eq:tildevbondi}
  \tilde V=1
\end{equation}
or equivalently
\begin{equation}
  \label{eq:dotalfa}
  \dot\alpha=V.
\end{equation}
With these relations the previous transformations are now
\begin{equation}\label{eq:tpsi2bondi}
\tilde \Psi_2^0 
= \frac{1}{V^3} 
\left( \Psi_2^0 - \frac{2\, \eth_V\, \alpha}{V} \Psi_3^0
+ \frac{(\eth_V \alpha)^2}{V^2}\, \Psi_4^0 \right) ,
\end{equation}
\begin{equation}\label{eq:tpsi1bondi}
\tilde \Psi_1^0 =  
\frac{1}{V^3} 
\left( 
\Psi_1^0 - 3 \frac{\eth_V\, \alpha}{V} \Psi_2^0 
+ 3 \frac{(\eth_V\, \alpha)^2}{V^2} \Psi_3^0
- \frac{(\eth_V \alpha)^3}{V^3}\, \Psi_4^0 
\right) ,
\end{equation}
\begin{equation}
\begin{split}
  \label{eq:tsigma0bondi}
  \tilde\sigma_0 =
 \frac{1}{V}
\left(\sigma_0 
+ \frac{1}{V} \eth_V^2 \alpha 
- \frac{1}{V^3 } \dot V \left(\eth_V \alpha \right)^2
  - \frac{2}{V^2} \eth_V \alpha \; \eth_V V
\right)
\end{split},
\end{equation}
\begin{equation}
  \label{eq:ethtildebondi}
  \tilde{\eth}_{\tilde V} f 
= \frac{1}{V}\eth_V f 
  - \frac{s}{V^2}  \eth_V V \;f
  -  \frac{1}{V^2} \eth_V \alpha \; \dot f
 .
\end{equation}

\subsection{Rest frame and center of mass conditions}\label{sec:center}

In order to study the condition for a section to be a rest frame
we need to consider the transformation properties of the supermomentum
integrand $\Psi$ defined in a Bondi system by
\begin{equation}
\Psi_B = 
\Psi_{B2}^0 + \sigma_{B0} \, \dot{\bar\sigma}_{B0} + \eth_B^2 \bar\sigma_{B0} 
;
\end{equation}
where the appearance of the subindex $B$ is used to 
emphasize that the quantities
refer to a Bondi system.

If $S$ is a nice section then, on $S$ one has
\begin{equation} \label{niceintrinsic}
\Psi_B = - K^3 M,
\end{equation}
where $K$ is given by equation (\ref{eq:K}). The idea is to consider the
original tetrad and coordinate system at scri $(u,\zeta,\bar\zeta)$ as coinciding
with a one parameter family of nice sections, defined by the conditions
$u={\tt constant}$.

The expression of the supermomentum psi $\Psi_B$ in terms of the intrinsic system 
is rather complicated, namely
\begin{equation}
  \label{eq:psibondi}
  \begin{split}
  \Psi_B = &
\Psi_2^0 \, V^{-3} - 2 \, (\bar\eth_V V)^2 \, \sigma_0 \, V^{-5} + \bar\eth_V^2
  V  \, \sigma_0 \, V^{-4} 
 + \bar\eth_V^2 \eth_V^2 \alpha \, V^{-4} \\
&- 4 \, \bar\eth_V \eth_V \alpha \, 
  \bar\eth_V \eth_V V  \, V^{-5} 
 - 2 \, \bar\eth_V \eth_V^2 \alpha \, \bar\eth_V V  \, V^{-5} - 2 \, \eth_V V  
  \, \bar\eth_V^2 \eth_V \alpha \, V^{-5} \\
& + 8 \, \eth_V V  \, \bar\eth_V \eth_V \alpha \, \bar\eth_V V  \, V^{-6} 
  - 2 \, (\eth_V V)^2 \, \bar\sigma_0 \, V^{-5} 
  - 2 \, \eth_V \alpha \, \bar\eth_V^2 \eth_V V \, V^{-5} \\
& + 8 \, \eth_V \alpha 
   \, \bar\eth_V \eth_V V  \, \bar\eth_V V  \, V^{-6} 
  - 8 \, \eth_V \alpha \, \eth_V V  \, (\bar\eth_V V)^2 \, V^{-7} + 2 
   \, \eth_V \alpha  \, \eth_V V \, \bar\eth_V^ V \, V^{-6} \\
& + \eth_V^2 V \, \bar\sigma_0 \, V^{-4} + \eth_V^2 \bar\sigma_0 \, V^{-3} 
  - \eth_V^2 \dot V \, (\bar\eth_V \alpha)^2 \, V^{-6} - 4 \, 
   \eth_V \dot V \, \bar\eth_V \eth_V \alpha \, \bar\eth_V \alpha \, V^{-6} \\
&  + 6 \, \eth_V \dot V \, \eth_V V \, (\bar\eth_V \alpha)^2 \, V^{-7} 
  + 4 
   \, \eth_V \dot V \, \eth_V \alpha \, \bar\eth_V \alpha \, \bar\eth_V V
   \, V^{-7} \\
&  + 2 \, \eth_V \ddot V  \, \eth_V \alpha \, (\bar\eth_V \alpha)^2 \, V^{-7} 
   - \dot V \, \sigma_0 \, \bar\sigma_0 \, V^{-4} 
  - 2 \, \dot V \, \bar\eth_V \alpha \, \bar\eth_V V \, \sigma_0 \, V^{-6} \\
&   - 2 \, \dot V \, (\bar\eth_V \eth_V \alpha)^2 \, V^{-6} 
  - 2 \, \dot V \, \bar\eth_V \eth_V^2 \alpha \, \bar\eth_V \alpha 
   \, V^{-6} + 12 \, \dot V \, \eth_V V \, \bar\eth_V \eth_V \alpha \, 
   \bar\eth_V \alpha \, V^{-7} \\
& - 10 \, \dot V \, (\eth_V V)^2 \, (\bar\eth_V \alpha)^2 \, V^{-8} + 2 
   \, \dot V \, \eth_V \alpha  \, \bar\eth_V \alpha \, V^{-4} 
  + 8 \, \dot V \, \eth_V \alpha \, \bar\eth_V \eth_V V \, 
   \bar\eth_V \alpha \, V^{-7} \\
&+ 4 \, \dot V \, \eth_V \alpha \, 
   \bar\eth_V \eth_V \alpha \, \bar\eth_V V \, V^{-7} 
  - 16 \, \dot V \, \eth_V \alpha \, \eth_V V \, \bar\eth_V \alpha 
   \, \bar\eth_V V \, V^{-8} \\
&  - \dot V \, (\eth_V \alpha)^2 \, 
   \bar\eth_V^2 V \, V^{-7} 
  - 2 \, \dot V \, (\eth_V \alpha)^2 \, \bar\eth_V \dot V \, 
   \bar\eth_V \alpha \, V^{-8} + 2 \, \dot V \, \eth_V^2 V \, 
   (\bar\eth_V \alpha)^2 \, V^{-7} \\
&  - 8 \, \dot V \, \eth_V \dot V \, \eth_V \alpha \, (\bar\eth_V \alpha)^2 
   \, V^{-8} + 2 \, \dot V^2 \, (\bar\eth_V \alpha)^2 \, \sigma_0 \, V^{-7} \\
& - 8 \, \dot V^2 \, \eth_V \alpha \, \bar\eth_V \eth_V \alpha \, 
   \bar\eth_V \alpha \, V^{-8} 
  + 16 \, \dot V^2 \, \eth_V \alpha \, \eth_V V \, (\bar\eth_V \alpha)^2 
   \, V^{-9} \\
& + \dot V^2 \, (\eth_V \alpha)^2 \, \bar\sigma_0 \, V^{-7} 
  + 12 \, \dot V^2 \, (\eth_V \alpha)^2 \, \bar\eth_V \alpha \, \bar\eth_V V \, V^{-9} \\
&  - 10 \, \dot V^3 \, (\eth_V \alpha)^2 \, (\bar\eth_V \alpha)^2 \, V^{-10} 
   - \ddot V \, (\bar\eth_V \alpha)^2 \, \sigma_0 \, V^{-6} 
  + 4 \, \ddot V \, \eth_V \alpha \, \bar\eth_V \eth_V \alpha \, 
   \bar\eth_V \alpha \, V^{-7} \\
& - 6 \, \ddot V \, \eth_V \alpha \, \eth_V V \, (\bar\eth_V \alpha)^2 \, V^{-8} 
   - 4 \, \ddot V \, (\eth_V \alpha)^2 \, \bar\eth_V \alpha \, \bar\eth_V V
   \, V^{-8} \\
&  + 8 \, \ddot V \, \dot V \, (\eth_V \alpha)^2 \, (\bar\eth_V \alpha)^2 
   \, V^{-9} 
  - \dddot V  \, (\eth_V \alpha)^2 \, (\bar\eth_V \alpha)^2 \, V^{-8} \\
&  + \dot{\bar\sigma}_0 \, \sigma_0 \, V^{-3} - \dot{\bar\sigma}_0 \, \dot V \, 
   (\eth_V \alpha)^2 \, V^{-6} .
  \end{split}
\end{equation}

However, it is important to note that the function $\alpha$ which generates
the transformation (\ref{eq:tildeu})-(\ref{eq:tildezeta}) is defined up to an
integration constant coming from the condition (\ref{eq:dotalfa}). Therefore
at each $u=u_0$ we can choose the integration constant such that $\alpha(u=u_0)=0$.
In other words, given $u_0$, one can always define
\begin{equation}
  \label{eq:alfau0}
  \alpha_{u_o}(u,\zeta,\bar\zeta) = \int_{u_0}^u V(u',\zeta,\bar\zeta)\, du' .
\end{equation}

Using this choice, equation (\ref{eq:psibondi}) simplifies considerably;
and the nice section equation becomes
\begin{equation}
  \label{eq:nicecondintrin}
    \begin{split}
 \Psi_B(u=u_0,\zeta,\bar\zeta) = &
\frac{1}{V^3}
\left(
  \Psi_2^0 + \dot{\bar\sigma}_0 \, \sigma_0 -\frac{\dot V }{V} \sigma_0 \, \bar\sigma_0
  + \eth_V^2 \bar\sigma_0
\right)
+
\frac{1}{V^4}
\left(
  \bar\eth_V^2 V \, \sigma_0 \, + \eth_V^2 V \, \bar\sigma_0 \, 
\right) \\
& -
\frac{2}{V^5}
\left(
   (\bar\eth_V V)^2 \, \sigma_0 + (\eth_V V)^2 \, \bar\sigma_0 
\right) \\
= & - K^3 \; M
,
    \end{split}
\end{equation}
where we are using as $\alpha$ the expression coming from (\ref{eq:alfau0}).

This constitutes the condition that must be satisfied at each section 
of a system to represent a one parameter family of nice sections.

The center of mass condition is equation (\ref{eq:acondit2}) which in the present 
notation means
\begin{equation}
\label{eq:centerofmasscond}
\begin{split}
Q_S(w)=  \int_{S}C   = 4\int_{S}
\Bigl(  - &\, w_{B2} \left( \Psi_{B1}^{0} 
+ 2\sigma_{B0} \eth_B\bar{\sigma}_{B0} +\eth_B\left( \sigma_{B0} \bar{\sigma}_{B0} \right)
\right)
\Bigr. \\
+ &\Bigl.\,
2\, w_{B1} \Psi_B
\Bigr) 
\;dS_B^{2} 
+ {\tt c.c.} = 0 
\end{split}
\end{equation}
for all $a=\bar a$ and $b=0$; and where we are already using the nice section 
condition.

All these quantities are evaluated with respect to the Bondi system, so that in
terms of the intrinsic system one must use the transformation laws of
equations (\ref{eq:tpsi2bondi})-(\ref{eq:ethtildebondi}).

\section{Final comments}\label{sec:final}

In this work we have shown how to circumvent the difficulties inherent in the 
structure of the BMS group by the use of the nice section construction; in
particular in the task of defining the intrinsic angular momentum in radiating
spacetimes. We have provided, as far as we know, with the only definition
of angular momentum which satisfies the expected physical properties, as those
discussed at the end of section \ref{sec:intrinsic}.

In particular, let us note that detectors of gravitational waves can be 
considered as observers at future null infinity. Our construction provides
for each point $p$ at scri with a prescription that singles out the
unique center of mass section containing $p$; and also with the appropriate
rest frame to calculate intrinsic quantities like angular momentum
and multipole moments; which are important for the description of the
gravitational waves that one wishes to detect.

It is worth to mention the relation of our work with previous presentations
of the notion of angular momentum; so next we briefly comment on some
(arbitrary, incomplete) selection of previous references.

Early works had the ideology suggested by the conservations laws due to 
Komar\cite{Komar59}; where associated with a symmetry $v^a$ one can
define the 2-form $A_{ab}=\nabla_{[a} v_{b]}$,
and define the `charge' $K(v)$\cite{Winicour80} on a 2-surface $S$ by
\begin{equation}\label{eq:Komar}
K_S(v) = -\frac{1}{8\pi}\int_S A^* ;
\end{equation}
where $A^*$ is the dual of $A$. This has the same form as the expression
that gives the electric charge $q$ of the electromagnetic field $F_{ab}$,
namely
\[
q (S) = -\frac{1}{8\pi}\int_S F^* .
\]
If $\Sigma$ is a hypersurface that has as boundary $S$, then from Stokes'
theorem one has
\[
K_S(v) = -\frac{1}{8\pi}\int_\Sigma dA^* ;
\]
the fact that this holds for any such hypersurface $\Sigma$, is sometimes
referred to as a conservation law.
Noting that $dA^*_{abc} = \frac{1}{3}\epsilon_{abcd}\nabla_e A^{ed}$,
one can see that if $v^a$ is a Killing symmetry then
$dA^*_{abc} = \frac{2}{3}\epsilon_{abcd} R_e^d v^e$ and therefore,
assuming Einstein equations,
the Komar charge can be related to the matter content of the spacetime.
If the sources are bounded by a world tube, then for any surface $S$
surrounding the world tube, the Komar charge will give the same value.
So, in the presence of a Killing symmetry, the Komar charge provides a
simple useful notion of a conserved quantity; in particular for
axis symmetric spacetimes, it provides with the notion of angular
momentum(one component). 

Although in a general spacetime one will not have a Killing symmetry,
the Winicour approach\-\cite{Tamburino66}\cite{Winicour68}\cite{Winicour80} 
uses the Komar charge idea, at future null infinity, associated to the 
asymptotic symmetries of an isolated system. When expressing the 
integral in terms of the asymptotic fields, there appears a term
with the factor $[2 \Psi_1^0 - 2\sigma_0 \eth \bar\sigma_0%
-\eth(\sigma_0 \bar\sigma_0)]$ plus another term containing the 
supermomentum part. One must first be aware that the curvature
spin coefficients definitions has been varying, and in some cases
it is even difficult to trace the conventions; in particular the relative
negative sign seems to be the consequence of difference with
our conventions\cite{Geroch73}; however, one can note
the non-trivial appearance of a factor of 2 in front of $\Psi_1^0$.
The Winicour charges were further developed in reference 
\cite{Geroch81} where a flux was calculated. The relation with
the symplectic fluxes was worked out in reference \cite{Ashtekar81}.

Using a gauge theory approach, Bramson\cite{Bramson75}, presents a notion
of angular momentum in which its integral has the factor
$[\Psi_1^0 - 2\sigma_0 \eth \bar\sigma_0-\eth(\sigma_0 \bar\sigma_0)]$
and where no term associated with supermomentum appears. In fact in
Bramson's approach no use of the generators of supertranslations is done.
As we have mentioned it is essential to make use of this term in 
order to be able to pick up the notion of intrinsic angular momentum.

Inspired by the Komar approach, Prior presented\cite{Prior77} a definition
of angular momentum that resembles the Winicour expressions but lacking
the supermomentum term.

Using a completely different approach, Streubel\cite{Streubel78}
found the same expression for the angular momentum as Bramson's,
but now with a supermomentum term.

Penrose presented\cite{Penrose82} a notion of quasi-local mass and angular 
momentum associated with any sphere in the interior of the spacetime;
making use of charge integrals of the Riemann tensor. His construction,
when extended to a sphere at future null infinity, gives an integral
expression where only curvature terms appear. However, as
Dray and Streubel have shown\cite{Dray84}, one can add the shear terms
with some arbitrary factors, due to some identities. They also
presented a charge associated to generators of an unfortunately
complex BMS Lie algebra.
Their choice is such that the angular momentum factor agrees with
Winicour's one.

Recent works\cite{Katz97}\cite{Wald00} have
concentrated on the relation between angular momentum
and Lagrangian formulations, although they use background geometries
in their constructions.

In reference \cite{Rizzi98} Rizzi presented a definition that tackles
the problem of supertranslation ambiguities. However, there is no mention in
this work of the issue of supermomentum, and therefore it is 
unable to set the center of mass frame; which is needed to define
the intrinsic angular momentum.

Independently from the nature of the approach all these definitions have
the characteristics that they provide with a notion of angular momentum
associated to a given Lorentz rotation BMS generator $v^a$ and a given
section $S$ of future null infinity. The choice of a Lorentz rotation
can be associated with a choice of a particular Bondi frame, because
to each Bondi frame there is a Lorentz subgroup of rotation that leave
the origin $(u=0)$ of the Bondi frame unchanged. Since among Bondi
frames there is the supertranslation freedom, there is an ambiguity
associated with this choice. But also any section $S$ can be thought
of as a section $u=\gamma(\zeta,\bar\zeta)$ of a given Bondi frame;
therefore there is also a supertranslation choice associated with
the election of section $S$. This becomes very important when one wants
to determine the intrinsic angular momentum of the sources.
All the definitions mentioned previously in this section suffer from this
supertranslations ambiguities.

In reference \cite{Moreschi86} we presented a definition that fixed this 
problem; that required a preferred choice of Lorentz generators based
on the properties of the spacetime in the limit for $u\rightarrow -\infty$;
that is in the limit to spacelike infinity. Therefore given any $S$
we provided with a charge integral of the Riemann tensor that had
the correct angular momentum behavior, with an immediate flux law and
without supertranslation ambiguities.  
The remaining difficulty with our previous work was that it would
not necessarily pick up the natural expected notion of intrinsic
angular momentum of the sources for the cases of radiating spacetimes.

From the naive comparison of the mathematical expression for angular momentum
factors on a given section $S$, one could distinguish mainly the following cases:


\begin{tabular}{|l|l|l|}
\hline 
Angular momentum factor & Supermomentum factor & appearing in references \\
\hline \hline
Winicour type & Winicour supermomentum & \cite{Tamburino66}\cite{Winicour68}\cite{Winicour80} \\
Winicour type & absent & \cite{Prior77} \\
Winicour type & Geroch supermomentum & \cite{Dray84}\cite{Wald00} \\
Winicour type & psi supermomentum & \cite{Katz97} \\
Bramson type & absent & \cite{Bramson75}\cite{Rizzi98} \\
Bramson type & psi supermomentum & \cite{Streubel78}\cite{Moreschi86} [present]\\
\hline
\end{tabular}



\noindent Actually Rizzi's integrand does not coincide with Bramson's exactly,
but it is closer to this than to Winicour type integrands.
The fact that two or more references appear in the same row, does not mean that 
they are equivalent. This table only shows the similarities of the factors
involving the curvature tensor; but the definitions are completed with the
factors involving the information of the BMS generator that one is using.
In particular, references \cite{Streubel78} and \cite{Moreschi86} both are 
inspired from Penrose work\cite{Penrose82}; but applying the definition of
reference \cite{Streubel78} to two sections $S_1$ and $S_2$, due to the 
supertranslation ambiguity problem, it is not clear how to relate the two
quantities.
Instead reference \cite{Moreschi86} constitutes an extension
of Penrose expression to the whole of scri, making sense for radiating 
spacetimes, solving the ambiguities of supertranslations by a 
physical  condition and providing with an immediate flux law, which is zero
in the absence of gravitational radiation. 
It should be stressed that the Penrose expressions can be
made to agree to Winicour or Bramson type angular momentum factors, equivalently
to agree with Geroch or psi supermomentum factors. It seems that in reference
\cite{Streubel78} the choice was intended to get the Bramson angular momentum
type, although later the author changed his mind for the Winicour type\cite{Dray84}; 
instead in reference \cite{Moreschi86} the stress was on the supermomentum
part, due to the properties of the psi 
supermomentum\cite{Moreschi86}\cite{Moreschi88}\cite{Moreschi98}\cite{Dain00'}, 
and as a consequence the angular momentum factor was obtained to be of Bramson type.

Our present construction; which can also be considered a development of the
Penrose approach, and a mayor improvement over reference \cite{Moreschi86}, 
provides for the first time with a global well behaved notion
of intrinsic angular momentum for radiating sources, and
also gives the notion of instantaneous center of mass frame; which is 
essential for the discussion of higher multipole moments. 

For example,
if one is studying a system whose gravitational radiation is 
appropriately described by the quadrupole radiation formula, at an 
instant of time; then in a long lapse of time, the corresponding
center of mass frame will be supertranslated with respect to the
original one. Therefore, in order to know what is the quadrupole
moment at a later time, one needs to settle what is
the center of mass frame first; which we have done here.

We next review some expected properties that one may require
a charge integral of BMS symmetries $v^a$ to satisfy. The first
\ref{item:last} are adapted from reference \cite{Ashtekar82} and we
have added another one.
\begin{enumerate}
\item $Q_S(v)$ should be linear in the BMS symmetry $v^a$. \label{item:first}
\item The expression $Q_S(v)$ should only involve local fields at $S$ on scri.
\item For the case when $v^a$ is a translation,  $Q_S(v)$ should agree
with the corresponding component of the Bondi momentum evaluated at $S$.
\item If  $v^a$ is the natural extension to scri of a Killing symmetry of the 
spacetime then  $Q_S(v)$ should be independent of $S$, if the vacuum
Einstein equations are satisfied en a neighborhood of scri, and 
it should be proportional to 
the Komar charge  $K_S(v)$ of equation (\ref{eq:Komar}).\label{it:killing}
\item In Minkowski space,  $Q_S(v)$ should vanish for all BMS symmetries $v^a$
and section $S$.
\item Given a symmetry $v^a$ there should exist a local flux $F(v)$, linear
in $v^a$, such that  $Q_S(v) -  Q_{S'}(v) = \int_\Sigma F(v)$; where 
$\Sigma$ is the region on scri bounded by $S$ and $S'$.
\item The flux $F(v)$ should vanish in absence of gravitational radiation.
\label{item:last}
\item Having three consecutive stages at scri, a first stationary region 
$\Sigma_1$, a second radiating stage $\Sigma_2$
and a third stationary region $\Sigma_3$, then 
there is a continuous prescription for the center of mass sections $S_{\tt cm}$
provided by the angular momentum calculated from $Q_S(v)$ such that
$J^a_b P^b=0$, where $J^{ab}$ is the angular momentum calculated
from $Q_{S_{\tt cm}}(v_{\tt cm})$, in which $v_{\tt cm}$ are the Lorentz rotations
leaving $S_{\tt cm}$ invariant, and $P^b$ is the Bondi momentum. \label{it:mejor}
\end{enumerate}

One can readily see that our definition of charge integrals of the Riemann tensor at 
scri, which uses the charge  $Q_{S}(v)$ of equation (\ref{eq:charge}), satisfies
all condition except \ref{it:killing}; although it is still an open question
whether condition \ref{it:killing} is satisfied or not. Ours is the only 
definition we know to satisfy property \ref{it:mejor}.

We have seen that the rest frame and center of mass condition for the intrinsic
systems are rather complicated. However one should have in mind that for
astrophysical interesting systems, the amount of gravitational radiation
is small, and therefore several quantities can be approximated around
stationary values.

This is precisely the key idea we plan to implement in future works.

\section*{Acknowledgments}
It is a pleasure to thank Dr. Dain for discussion on these topics
and Dr. Rizzi for ongoing discussion on his work.
We acknowledge support from SeCyT-UNC and CONICET.



\end{document}